\documentstyle[11pt]{article}

\begin{document}

\begin{center}

{\Large{\bf Dynamic Critical Phenomena in Trapped Bose Gases}} \\ [5mm]

{\large V.I. Yukalov$^1$, E.P. Yukalova$^2$ and V.S. Bagnato$^3$} \\ [3mm]

$^1$Bogolubov Laboratory of Theoretical Physics\\
Joint Institute for Nuclear Research, Dubna 141980, Russia \\ [3mm]

$^2$Department of Computational Physics\\
Laboratory of Informational Technologies\\
Joint Institute for Nuclear Research, Dubna 141980, Russia \\ [3mm]

$^3$Instituto de Fisica de S\~ao Carlos, Universidade de S\~ao Paulo \\
Caixa Postal 369, S\~ao Carlos, S\~ao Paulo 13560-970, Brazil

\end{center}

\vskip 1cm

\begin{center}

{\bf ABSTRACT}

\end{center}

{\parindent=0pt

Nonlinear dynamics of a trapped Bose-Einstein condensate, subject to 
the action of a resonant external field, is studied. This field produces a 
spatio-temporal modulation of the trapping potential with the frequency 
close to the transition frequency between the ground state and a higher 
energy level. The evolution equations of fractional populations display a 
kind of critical phenomena at a critical line on the manifold of the system    
parameters. It is demonstrated that there exists a direct analogy between 
dynamical instability at this line and critical phenomena at a critical line 
of a related averaged system. 

\vskip 3mm

}

\vskip 1cm

A dilute cloud of trapped Bose-condensed atoms, at low temperatures, can be 
described by a wave function satisfying the Gross-Pitaevskii equation [1,2]. 
This equation is nonlinear due to the atomic interactions through a local 
delta potential. The mathematical structure of the equation is that of the 
nonlinear Schr\"odinger equation. Stationary states of the latter, because of 
the confinement caused by a trapping potential, are restricted to discrete 
energy levels. These stationary eigenstates form a set of solutions that can 
be called {\it nonlinear modes}, in analogy to linear modes that are solutions 
of a linear Schr\"odinger equation. The nonlinear modes can also be called 
{\it coherent modes}, since the wave function of the Gross-Pitaevskii equation 
corresponds to a coherent state of Bose-condensed atoms. One can term as well 
these modes as {\it topological modes}, emphasizing that the wave functions 
related to different energy levels have different spatial topology.

The modal structure of the confined condensate states is in close analogy 
with {\it nonlinear optical modes} of nonlinear waveguide equations in 
optics [3,4]. And the methods of creating nonlinear coherent modes of trapped
Bose atoms [5,6] are also similar to those employed in optics, where one 
uses specially prepared initial conditions or invokes an action of external 
fields. Dynamics of fractional populations, characterizing the occupation
of coherent modes of Bose-condensed trapped atoms, displays as well many 
phenomena equivalent to those known in optics, for instance collapses, 
revivals, and Rabi-type oscillations. The nonlinear dynamics of processes 
coupling bosonic coherent modes have been studied in several papers 
considering the general case [5,6], antisymmetric modes [7,8], dipole 
topological modes in a two-component condensate [9,10], and vortex modes 
[11-14].

The aim of the present paper is twofold: First, we demonstrate the existence 
of a critical line for the population dynamics of Bose condensates, which is 
due to the nonlinearity of evolution equations. Second, we show that there is 
an intimate relation between the studied nonlinear dynamical system and a 
related averaged system, so that the instabilities happening at a critical 
line of the dynamical system are the counterparts of critical phenomena 
occurring in the averaged system.

The time-dependent Gross-Pitaevskii equation has the form
\begin{equation}
\label{1}
i\hbar\;\frac{\partial\varphi}{\partial t} = \left [ \hat H(\varphi) +
\hat V \right ]\; \varphi\; , \qquad
\hat H(\varphi)= \; - \; \frac{\hbar^2}{2m_0}\; {\vec\nabla}^2 +
U(\vec r\;) + AN|\varphi|^2 \; ,
\end{equation}
where $U(\vec r)$ is a trapping potential; $A\equiv4\pi\hbar^2a_s/m_0$; 
$a_s$ is a scattering length; $m_0$ is mass; and $N$ is the number of 
particles; $\hat V$ is a potential of external fields. The wave function 
$\varphi$ is normalized to unity, $||\varphi||=1$. The nonlinear 
topological modes are defined as the eigenfunctions of the nonlinear 
Hamiltonian, that is, they are solutions to the eigenproblem
$$
\hat H(\varphi_n)\varphi_n(\vec r) =E_n\varphi_n(\vec r).
$$ 

It is worth stressing here the principal difference between the topological 
coherent modes and elementary collective excitations. The former are 
self-consistent {\it atomic} states defined by the {\it nonlinear} 
Gross-Pitaevskii equation. While the latter are the {\it elementary} 
excitations corresponding to small deviations from a given atomic state and 
described by the {\it linear} Bogolubov-De Gennes equations [1,2].

We assume that at the initial time the system was condensed to the 
ground-state level with an energy $E_0$. So that the initial condition to 
eq. (1) is $\varphi(\vec r,0)=\varphi_0(\vec r)$. Suppose, we wish to couple 
the ground state $\varphi_0$ with another state $\varphi_j$ having a higher 
energy $E_j$. The best way for doing this, is clearly, by switching on an 
external field, say $\hat V= V(\vec r)\cos\omega t$, oscillating with a 
frequency $\omega$ which is close to the transition frequency 
$$
\omega_j\equiv(E_j-E_0)/\hbar,
$$
so that the detuning $\Delta\omega\equiv\omega-\omega_j$ be small, 
$|\Delta\omega/\omega|\ll 1$. Then one can look for a solution to eq. (1) in 
the form
\begin{equation}
\label{2}
\varphi(\vec r,t) = c_0(t)\varphi_0(\vec r)\; e^{-iE_0t/\hbar} +
c_j(t)\varphi_j(\vec r)\; e^{-iE_jt/\hbar} \; .
\end{equation}
The considered situation is analogous to the description of nonlinear 
resonant processes in optics [3,4]. The validity of the quasiresonant 
two-level approximation (2) for the time-dependent Gross-Pitaevskii equation 
has been confirmed from the general point of view [5,6] and also proved for 
several similar cases by direct numerical simulations of the Gross-Pitaevskii 
equation [8,9,11-13], the agreement between the two-level picture and the 
simulations being excellent.

The coefficients $c_i(t)$ define the fractional level populations 
$n_i(t)\equiv |c_i(t)|^2$. The equations for these coefficients can be 
obtained by substituting the presentation (2) in eq. (1). It is again worth 
noticing the similarity of this presentation with the slowly-varying-amplitude 
approximation commonly used in optics, if one treats the factors $c_i(t)$ as 
slow functions of time, as compared to the exponentials in eq. (2), which 
implies that $|dc_i/dt|\ll E_i$. This approximation, being complimented by
the averaging technique [15], permits one to slightly simplify the evolution 
equations for $c_i(t)$. In this way [5,6], we come to the equations
\begin{equation}
\label{3}
\frac{dc_0}{dt} =\; - i\alpha\; n_jc_0 -\; \frac{i}{2}\; \beta\; c_j\;
e^{i\Delta\omega t} \; , \qquad
\frac{dc_j}{dt} =\; - i\alpha\; n_0c_j -\; \frac{i}{2}\; \beta^*\; c_0 \;
e^{-i\Delta\omega t}\; ,
\end{equation}
in which the transition amplitudes
$$
\alpha_{ij} \equiv A\;\frac{N}{\hbar}\; \int |\varphi_i(\vec r)|^2
\left ( 2|\varphi_j(\vec r)|^2 -|\varphi_i(\vec r)|^2\right ) \; d\vec r\; ,
\qquad
\beta \equiv\; \frac{1}{\hbar}\; \int \varphi_0^*(\vec r) V(\vec r)
\varphi_j(\vec r)\; d\vec r \; ,
$$
caused by the nonlinearity and by the modulating field, respectively, are 
introduced, and the abbreviated notation $\alpha\equiv\alpha_{0j}$, with 
setting, for simplicity, $\alpha_{0j}=\alpha_{j0}$, is used. Note that the 
transition amplitude $\beta$ is nonzero only if the potential $V(\vec r)$ 
depends on space variables. The initial conditions to eqs. (3) are $c_0(0)=1$ 
and $c_j(0)=0$.  

Usually, one solves such evolution equations with fixed parameters $\alpha,\;
\beta$ and $\Delta\omega$, keeping in mind a particular experimental 
realization. Instead of this, we have studied the behavior of solutions to 
eqs. (3) in a wide range of varying parameters. It turned out that this 
behaviour is surprisingly rich exhibiting new interesting effects.

First of all, it is easy to notice that the number of free parameters in 
eqs. (3) can be reduced to two by the appropriate scaling. For this purpose, 
we measure time in units of $\alpha^{-1}$ and introduce the dimensionless 
parameters 
$$
b\equiv|\beta|/\alpha, \qquad  \delta\equiv\Delta \omega/\alpha.
$$
It is also evident that eqs. (3) are invariant with respect to the inversion  
$$
\alpha\rightarrow -\alpha, \qquad  \beta\rightarrow -\beta, \qquad 
\Delta\omega\rightarrow -\Delta\omega, \qquad  t\rightarrow -t.
$$ 
Therefore it is possible to fix the sign of one of the parameters, say 
$\alpha>0$, since the opposite case can be obtained by the inversion. For 
concreteness, we shall also keep in mind that $\beta$ is positive. The 
dimensionless detuning is assumed to always be small, $|\delta|\ll 1$. And 
the dimensionless transition amplitude $b$ is varied in the region 
$0\leq b\leq 1$. We have accomplished a careful analysis by numerically 
solving eqs. (3). When parameter $b$ is small, the fractional populations 
oscillate reminding the Rabi oscillations in optics, where $|\beta|$ would 
play the role of the Rabi frequency. The amplitude of these oscillations 
increases with increasing $b$. It would be more correct to say that in our 
case there exist {\it nonlinear Rabi oscillations}, as far as eqs. (3) differ 
from the corresponding equations for optical two-level systems by the 
presence of the nonlinearity due to interatomic interactions. This 
nonlinearity not only slightly modifies the Rabi-type oscillations of the 
fractional populations but, for a particular relation between parameters, 
can lead to dramatic effects. By accurately analyzing the behaviour of 
solutions to eqs. (3), with gradually varying parameters, we have found out 
that there exists the {\it critical line} 
$$
b+\delta\simeq 0.5,
$$ 
at which the system dynamics experiences sharp changes [6]. For example, if 
the parameter $b=0.4999$ is kept fixed, and the critical line is crossed by 
varying the detuning $\delta$, the following phenomena occur. When the 
detuning is zero, the fractional populations display the Rabi-type 
oscillations. After slightly shifting the detuning to $\delta=0.0001$ the top 
of $n_j(t)$ and the bottom of $n_0(t)$ become flat, and the oscillation 
period is approximately doubled. A tiny further variation of the detuning to 
$\delta=0.0001001$ yields again drastic changes, so that the upward cusps of 
$n_j(t)$ and the downward cusps of $n_0(t)$ arise. The following small 
increase of the detuning to $\delta=0.00011$ squeezes the oscillation period 
twice. Then, making $\delta$ larger does not result in essential qualitative 
changes of the population behaviour. All dramatic changes in dynamics occur 
in a tiny vicinity of the critical line. The same phenomena happen when 
crossing the line $b+\delta\simeq 0.5$ at other values of parameters or if 
$\delta$ is fixed but $b$ is varied.

The unusual behaviour of the fractional populations is due to the nonlinearity
of the evolution equations (3). Systems of nonlinear differential equations, 
as is known, can possess qualitatively different solutions for parameters 
differing by infinitesimally small values. The transfer from one type of 
solutions to another type, in the theory of dynamical systems, is, generally, 
termed bifurcation. At a bifurcation line, dynamical system is structurally 
unstable.

The second aim of our paper is to show that the found instability in the
considered dynamical system is analogous to a {\it phase transition} in a
statistical system. To elucidate this analogy for the present case, we have 
to consider the time-averaged features of the dynamical system given by 
eqs. (3). To this end, we need, first, to define an effective Hamiltonian 
generating the evolution equations (3). This can be done by transforming 
these equations to the Hamiltonian form 
$$
i dc_0/dt=\partial H_{eff}/\partial c_0^*\; , \qquad     
i dc_j/dt=\partial H_{eff}/\partial c_j^*, 
$$
with the effective Hamiltonian 
\begin{equation}
\label{4}
H_{eff} = \alpha\; n_0n_j + \frac{1}{2}\left ( \beta\; e^{i\Delta\omega t}
c_0^*c_j + \beta^*\; e^{-i\Delta\omega t} c_j^* c_0\right ) \; .
\end{equation}
An effective energy of the system can be defined as a time average of the 
effective Hamiltonian (4). For this purpose, eqs. (3) can be treated by 
means of the averaging technique [15], as it is described in detail in 
Ref. [5], which provides the guiding-center solutions. Substituting the 
latter in eq. (4), together with the time-averaged fractional populations, 
results in the effective energy 
$$
E_{eff} =(\alpha b^2/2\varepsilon^2)(b^2/2\varepsilon^2+\delta),
$$
where $\varepsilon$ is a dimensionless average frequency defined by the 
equation 
$$
\varepsilon^4(\varepsilon^2-b^2)=(\varepsilon^2-b^2-\varepsilon^2\delta)^2.
$$
This effective energy, represents the time average of the effective 
Hamiltonian (4) and, thus, characterizes the average features of the system. 
As an {\it order parameter} for this averaged system, one can take the 
difference of the time-averaged populations, 
$$
\eta\equiv\overline n_0 -\overline n_j=1-b^2/\varepsilon^2.
$$ 
The capacity of the system to store the energy pumped in by the resonant 
field can be described by the {\it pumping capacity} 
$$
C_\beta=\partial E_{eff}/\partial|\beta|.
$$
The influence of the detuning on the order parameter is characterized by 
the {\it detuning susceptibility} 
$$
\chi_\delta=|\partial\eta/\partial\delta|.
$$ 
Analyzing the behaviour of the introduced 
characteristics as functions of the parameters $b$ and $\delta$, we found 
out that they exhibit critical phenomena at the {\it critical line} $b+\delta=
0.5$, which coincides with the bifurcation line for the dynamical system. 
Expanding these characteristics over the small relative deviation 
$\tau\equiv|b-b_c|/b_c$ from the {\it critical point} $b_c=0.5-\delta$, we 
obtain 
\begin{equation}
\label{5}
\eta-\eta_c\simeq\frac{\sqrt{2}}{2}( 1 - 2\delta)\tau^{1/2}\; , \qquad
C_\beta\simeq\frac{\sqrt{2}}{8}\tau^{-1/2}\; , \qquad
\chi_\delta\simeq\frac{1}{\sqrt{2}}\tau^{-1/2}\; ,
\end{equation} 
where $\eta_c\equiv\eta(b_c)$ and $\tau\rightarrow 0$. As is seen, the 
pumping capacity and detuning susceptibility display divergence at the 
critical point. The related {\it critical indices} for $C_\beta,\; \eta$, 
and $\chi_\delta$ are equal to $1/2$. These indices satisfy the known scaling 
relation: 
$$
ind(C_\beta)+2\; ind(\eta)+ind(\chi_\delta)=2,
$$ 
where $ind$ is the evident abbreviation for index.

In order to clarify what is the origin of the found critical effects for 
the studied dynamical system, let us return back to eqs. (3). Again we pass 
to dimensionless notation measuring time in units of $\alpha$. By means of 
the substitution
\begin{equation}
\label{6}
c_0 = \left ( \frac{1-p}{2}\right )^{1/2} \; \exp \left\{ i\left (
q_0 + \frac{\delta}{2}\; t\right ) \right \} \; , \qquad
c_j = \left ( \frac{1+p}{2}\right )^{1/2} \; \exp \left\{ i\left (
q_1 -\; \frac{\delta}{2}\; t\right ) \right \} \; , 
\end{equation}
where $p,\; q_0$, and $q_1$ are real functions of $t$, equations (3) can be 
reduced to the form
\begin{equation}
\label{7}
\frac{dp}{dt} = - b\; \sqrt{1-p^2}\; \sin q \; , \qquad
\frac{dq}{dt} = p + \frac{bp}{\sqrt{1-p^2}}\; \cos q + \delta \; ,
\end{equation}
in which $q\equiv q_1-q_0$. Note that this reduction to an autonomous 
dynamical system is valid for arbitrary detuning $\delta$. Moreover, this 
system possesses the integral of motion  
$$
I(p,q)=\frac{1}{2}p^2-b\sqrt{1-p^2}\cos q +\delta p,
$$ 
which can be defined by using the initial conditions $p(0)=-1$ and $q(0)=0$ 
corresponding to the conditions $c_0(0)=1$ and $c_j(0)=0$. This gives 
$$
I(-1,0)=\frac{1}{2}-\delta.
$$ 
The existence of the integral of motion means that the dynamical system is 
integrable in quadratures. This fact does not help much for studying the time 
evolution of the system, since the formal solutions $p(t)$ and $q(t)$ are 
expressed through rather complicated integrals, so that the system evolution, 
anyway, is to be analysed numerically. However, the property of 
integrability implies that the appearance of chaos in the system is 
impossible. Consequently, the observed critical effects in no way could be  
related to chaos. Then what is their origin? The answer to this question
comes from the analysis of the phase portrait for eqs. (7) in the rectangle  
defined by the inequalities  
$$
-1\leq p\leq 1, \qquad  0\leq q\leq  2\pi.
$$ 
This analysis shows that, if $b+\delta<0.5$, then the motion starting at the 
initial point $p(0)=-1$, $q(0)=0$ is oscillatory, with a trajectory lying 
always in the lower part of the phase rectangle, below the separatrix given 
by the equation  
$$
\frac{1}{2}p^2-b\sqrt{1-p^2}\cos q +\delta p -\beta=0.
$$ 
When $b+\delta=0.5$, the separatrix touches the initial point, so that the 
following motion occurs in the phase region above the separatrix. In this 
way, if we consider, under the given initial conditions, the parametric 
manifold formed by the parameters $b\in[0,1]$ and $\delta\ll 1$, then the 
critical line $b+\delta=0.5$ separates the parametric regions related to 
two different types of solutions to eqs. (7).

Summarizing, we have considered the population dynamics of a trapped 
Bose-Einstein condensate, subject to the action of a resonant 
spatio-temporal modulation of the trapping potential. The consideration has 
been reduced to a two-level picture corresponding to the quasiresonant  
approximation. A careful analysis of the derived two-level evolution 
equations has been made for the wide range of varying parameters. Such a 
variation of parameters can be easily realized by changing trap 
characteristics, varying the number and kind of condensed atoms, and by 
changing the scattering lengths using Feshbach resonances [1,2]. It turned 
out that on the manifold of possible parameters there exists a critical line,  
where the evolution equations display structural instability. We have 
demonstrated that this instability for a dynamical system is analogous to a 
phase transition for a stationary averaged system. For the latter, one can 
define an order parameter, pumping capacity, and detuning susceptibility 
which exhibit critical phenomena at the critical line. The origin of this 
critical line is elucidated by showing that it divides the parametric 
manifold onto two regions corresponding to qualitatively different solutions 
of the evolution equations.

The consideration of the nonlinear dynamics of trapped Bose-condensed gases 
can be based on the two-level approximation because of the resonant 
character of the modulating external field. However, the accuracy of this 
approximation remains undetermined. A complete analysis of the dynamics would 
require the numerical solution of the initial Gross-Pitaevskii equation. Such 
a numerical investigation of the problem should give definite answers to the 
following questions:

(i) What are the conditions when the effect of the resonant excitation of 
coherent topological modes becomes possible?

(ii) What types of coherent modes could be actually excited by this method? 
Could we create vortex modes in this way? 
     
(iii) What is the accuracy of the two-level approximation in the case of a 
resonant modulating field?

(iv) What is an exact equation for the critical line, where dynamic critical 
phenomena may appear? 

(v) More generally, what are the critical manifolds in the parametric space, 
when there are several system parameters, and what is the relation of these 
manifolds to the bifurcation manifolds?

Answering these questions would provide practical suggestions for creating 
various coherent modes of atom lasers. The corresponding numerical 
investigation is now in progress.

\vskip 5mm

{\bf Acknowledgement}

\vskip 3mm

One of us (V.I.Y.) is very grateful to V.K. Melnikov  for several useful 
discussions.

\end{document}